\documentclass[twocolumn,twoside,slac_two]{revtex4}
\usepackage{graphicx}
\usepackage{fancyhdr}
\newcommand{\mytilde}{\raise.17ex\hbox{$\scriptstyle\mathtt{\sim}$}}

\begin{document}

\title{New Methods for Timing Analysis of Transient Events, Applied to Fermi/GBM Magnetar Bursts}

%
\author{Daniela Huppenkothen, Anna L. Watts, Phil Uttley,  Alexander J. van der Horst, Michiel van der Klis}
\affiliation{Astronomical Institute ``Anton Pannekoek'', University of
  Amsterdam, Postbus 94249, 1090 GE Amsterdam, the Netherlands}

\author{Chryssa Kouveliotou}
 \affiliation{Office of Science and Technology, ZP12, NASA Marshall Space Flight Center, Huntsville, AL 35812, USA}

\author{Ersin G{\"o}{\u g}{\"u}{\c s}}
\affiliation{Sabanc\i~University, Orhanl\i-Tuzla, \.Istanbul  34956, Turkey}

\author{Jonathan Granot} 
\affiliation{The Open University of Israel, 1 University Road, POB 808, Ra’anana 43537, Israel}

\author{Simon Vaughan} 
\affiliation{X-Ray and Observational Astronomy Group, University of Leicester, Leicester, LE1 7RH, UK}

\author{Mark H. Finger}
\affiliation{Universities Space Research Association, Huntsville, AL 35805, USA}

\begin{abstract}
In order to discern the physical nature of many gamma-ray sources in the sky, we must look not only in spectral and spatial dimensions, but also understand their temporal variability. However, timing analysis of sources with a highly transient nature, such as magnetar bursts, is difficult: standard Fourier techniques developed for long-term variability generally observed, for example, from AGN often do not apply. Here, we present newly developed timing methods applicable to transient events of all kinds, and show their successful application to magnetar bursts observed with Fermi/{\it GBM}. Magnetars are a prime subject for timing studies, thanks to the detection of quasi-periodicities in magnetar Giant Flares and their potential to help shed light on the structure of neutron stars. 
Using state-of-the art statistical techniques, we search for quasi-periodicities (QPOs) in a sample of bursts from Soft Gamma Repeater SGR J0501+4516 observed with Fermi/{\it GBM} and provide upper limits for potential QPO detections. Additionally, for the first time, we characterise the broadband variability behaviour of magnetar bursts and highlight how this new information could provide us with another way to probe these mysterious objects. \end{abstract}

\maketitle

\thispagestyle{fancy}


\section{Introduction}
Neutron stars present the best test cases for extreme physics in the high-density regime.  A long-standing problem in neutron star physics is our lack of understanding of the neutron star interior, in particular, the dense matter equation of state \citep{Lattimer07}. The detection of quasi-periodic oscillations (QPOs) in the tails of giant flares from Soft Gamma Repeaters (SGRs) has opened up the possibility of studying neutron star interiors using asteroseismology (see \citealt{Watts11} for a review).  

SGRs exhibit regular bursts in the hard X-rays and soft $\gamma$-rays ($\lesssim 100 \, \mathrm{keV}$), and very rare giant flares with extremely high isotropic equivalent radiated energy of up to $10^{46} \, \mathrm{erg}$ (see e.g. \citealt{2005Natur.434.1107P}). Observations of persistent soft X-ray counterparts showing coherent pulsations with large periods of $5-8$ seconds \citep{1998Natur.393..235K, 1999ApJ...510L.115K}, and the detection of the same periodicities in the tails of the giant flares \citep{1999Natur.397...41H, 2005Natur.434.1107P}, suggested that SGRs are neutron stars.  Their behavior is understood within the context of the magnetar model \citep{1995MNRAS.275..255T}:  in this paradigm the SGRs are isolated neutron stars with exceptionally strong external dipole magnetic fields, with internal fields that may be as high as\footnote{Supported by period and period derivative measurements; see $\mathrm{http://www.physics.mcgill.ca/}$\textasciitilde$\mathrm{pulsar/magnetar/main.html}$ for an up-to-date reference list on magnetar spin-down properties} $10^{16} \, \mathrm{G}$. Giant flares are powered by a catastrophic reordering of the magnetic field \citep{2001ApJ...552..748W}.  Since this field is coupled to the solid crust, \citet{1998ApJ...498L..45D} suggested that such large-scale reconfiguration might rupture the crust, creating global seismic vibrations that would be visible as periodic modulations of the X-ray and $\gamma$-ray flux. This idea was confirmed by the detection of QPOs in the expected range of frequencies  ($\sim 10-1000$ Hz) in the tails of giant flares from two different magnetars \citep{2005ApJ...628L..53I, 2005ApJ...632L.111S, 2006ApJ...653..593S, 2006ApJ...637L.117W}. 
If the QPO frequencies can be reliably identified with particular global seismic modes of the neutron star, then they can in principle be used to constrain both the equation of state and the interior magnetic field.     

A major obstruction to this field of research is the sparsity of data. Since the launch of the first X-ray and $\gamma$-ray instruments, only three giant flares have been observed, with just two having data with a sufficient time resolution to detect QPOs. In trying to overcome this lack of observational constraints, it is therefore a reasonable approach to turn to the much more numerous short SGR bursts with lengths of usually less than a second and luminosities around $10^{40} \, \mathrm{erg}\, \mathrm{s}^{-1}$. Hundreds of SGR bursts have now been observed from many magnetars\footnote{see e.g. \citealt{woods06}, \citealt{mereghetti08} for overviews or http://f64.nsstc.nasa.gov/gbm/science/magnetars/ for a collection of SGR bursts observed with Fermi/{\it GBM}}. 

To date there has been no systematic search for periodic features in the lightcurves of the SGR bursts.  A search for QPOs in a period of enhanced emission with multiple bursts (a `burst storm'), from the magnetar SGR J1550-5418, carried out using data from the Fermi Gamma-ray Burst Monitor (GBM), found no significant signals \citep{Kaneko10}. \citet{2010ApJ...721L.121E} searched a subset of Rossi X-ray Timing Explorer data from SGR 1806-20 for periodic features and found some tentative signals: however, there are several points of concern with regard to their methodology explained in full in the Appendix of \citet{sgr0501_timing}.   

Searching for QPOs in transient light curves is a non-trivial task. Standard methods involve Fourier analysis, more specifically the periodogram, defined as the squared amplitude of the Fourier transform of the light curve. The periodogram has several advantageous statistical properties: Poisson noise prevalent in light curves from photon counting experiments results in a flat periodogram with a well-known statistical distribution, making the detection of periodic and quasi-periodic features (as outliers above that distribution) tractable.
However, the detectability of QPOs changes significantly in the presence of correlated noise processes and the transient properties of the bursts we are concerned with here.
The very nature of a transient event - it has a start, one or more peaks, and an end - complicates the analysis procedure and introduces additional sources of uncertainty, especially in the low-frequency part of the periodogram. For transient events where the shape of the burst envelope is known, many problems arising from the non-stationarity can be solved either analytically \citep{2011MNRAS.415.3561G} or via Monte Carlo simulations \citep{2001MNRAS.321..776F}. However, many astrophysical transients such as magnetar bursts do not show a well-behaved burst light curve that is easily reproducible by a simple function. This in itself can be interesting, aside from searching for QPOs: the different burst envelope shapes must be created by a physical process in the source, either in the form of noise processes or non-stochastic emission processes, and characterising the differences may tell us more about the emission processes at work. 
The methods and analysis presented in the following summarise more extensive work laid out in \citet{sgr0501_timing}. We refer the reader to that paper for more details.

\section{Variability Analysis}
\subsection{Monte Carlo Simulations of Light Curves - Advantages and Shortcomings}
Monte Carlo simulations of light curves are a standard tool in timing analysis (see for example \citealt{2001MNRAS.321..776F}). The underlying idea is simple: one fits an empirically derived (or physically motivated) function to the burst light curve. One then generates a large number of realizations of that burst profile, including appropriate sources of noise, such as Poisson photon counting noise. 
The periodograms computed from these fake light curves form a basis against which to compare the periodogram of the real data. For each frequency bin, a distribution of powers is produced.
Comparing the observed power in each bin with the distribution of simulated powers in the same bin allows us to make a statement about the probability of the observed power in a particular bin being due to a noise process: if the observed power in a particular bin is a high outlier compared with the distribution of simulated powers in that bin, then the probability of observing the data under the (null) hypothesis of a noise process is $1/N$, where $N$ is the number of simulations performed. If $N$ is large, the observed outlier is unlikely to be produced by the noise process alone. 

The Monte Carlo method outlined above is versatile and powerful, but it has limitations. The most important limitation comes from our lack of knowledge of the underpinning physical processes producing the observed light curve. Only if the null hypothesis accurately reflects the data - apart from the (quasi-)periodic signal for which we would like to test - is the test meaningful. If important effects that distort either shape or distribution of the powers are missed, then the predictions made will not be accurate, leading to either spurious detections or real signals not being found.

More often than not, especially in the case of short magnetar bursts, we do not have complete information about the emission mechanism. Short magnetar bursts are extremely diverse, varying in light curve shape as well as burst intensity and duration (see, for example, \citealt{1999ApJ...526L..93G} and  \citealt{2000ApJ...532L.121G}). There is a fundamental degeneracy in the problem: which features to we consider to be part of the burst envelope, and which a (potential) feature which we do not include in the light curve template? Stochastic processes correlated on different time scales can mimic a quasi-periodic process to the human eye, but clearly, this is not what we wish to detect.
Not taking the presence of these features into account can lead to a large number of false positive detections. On the other hand, overcorrecting for features may lead us to detect no features at all. 
Without detailed knowledge of the burst emission processes, it is impossible to build a reliable model for individual burst light curves to test agains when searching for QPOs. We therefore advocate a different method, based on an empirical model of the periodogram. 

\begin{figure*}
\begin{center}
\includegraphics[width=11.5cm]{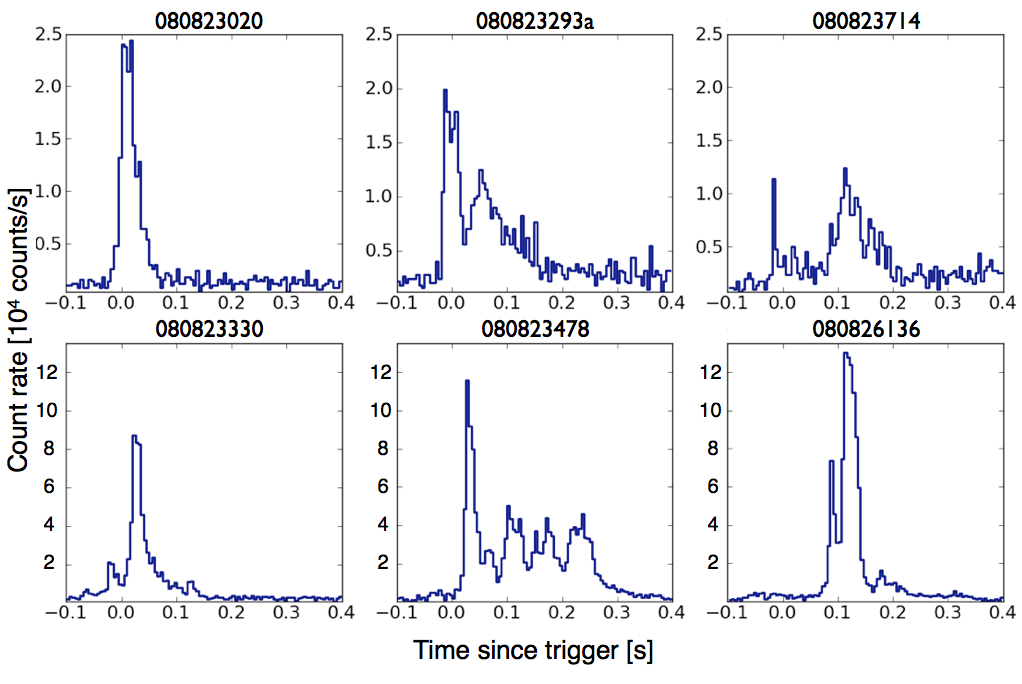}
\caption{Light curves of six example bursts from the magnetar SGR J$0501+4516$ recorded by Fermi/{\it GBM}. We combined data from all NaI detectors with source angles smaller than $50$ degrees to the source. The time resolution corresponds to $0.005$ seconds. Note the strong component of aperiodic variability after the main burst in $080823478$ and the differences in peak count rate by almost one order of magnitude between the upper three bursts and the lower three.}
\label{fig:burstselect}
\end{center}
\end{figure*}

\subsection{Modeling the Periodogram}

We give a very short overview of the general principles our method employs. Details can be found in \citet{sgr0501_timing} and \citet{vaughan2010}
The Bayesian procedure we employ in modeling the periodogram has three parts: (a) find the preferred broadband noise model to represent the low-frequency part of the periodogram, (b) search the periodogram for the highest outlier and compare this outlier to those distributed by pure broadband noise to find narrow features, (c) search for QPOs in the data, using binned data as well as an identical approach for the model selection in the first step. A step-by-step description can be found in the Appendix of \citet{sgr0501_timing}.

Every step in the analysis follows the same logic: assume a null hypothesis and an alternative hypothesis, compute statistics to summarise the data-model fits for the two different models, generate a sample from this null hypothesis using a Markov Chain Monte Carlo (MCMC) approach, more specifically the MCMC code {\it emcee} \citep{2012arXiv1202.3665F}. One can then compare the distribution of the relevant statistic derived from the sample generated from the null hypothesis to the observed value of that statistic.
If the observed value lies in the high-end tail of the distribution, then it is an outlier with respect to the null hypothesis. 
 
Since the entire procedure rests on the correct choice of broadband model, this is the first step of the analysis. The data are fitted with two continuum noise models, which, by definition of the likelihood ratio test, are required to be nested. The likelihood ratio is the statistic we use to decide which model is preferred by the data. We simulate a large number of fake periodograms from parameter sets drawn from the posterior distribution of parameters, as approximated by a large number of MCMC simulations. Then these fake periodograms are fit with both models again to build a distribution of likelihood ratios from the simple model. We can compute the tail-area probability (p-value) of the observed likelihood ratio to be typical of the distribution (equivalent to asking whether the observed data is sufficiently described by the simpler model) by integrating over the tail of the distribution. If this probability is lower than a chosen significance threshold, then the data is more likely to be drawn from the more complex model hypothesis, which should then be adopted for the rest of the analysis. 

We extensively tested our method on synthetic data generated with known parameters. These simulations confirm that in the limit of white noise, our method matches the predictions from standard Fourier analysis. Furthermore, for more complicated light curves, involving red noise and a burst envelope, our method is conservative in nature at low frequencies, where burst envelope and red noise dominate, but approaches the white noise predictions for high frequencies. The full description of these simulations is given in \citet{sgr0501_timing}.

\section{Observations}

To test our methodology, we applied it to a small sample of bursts from SGR J$0501+4516$. Fermi/{\it GBM} triggered 26 times on this source between 2008 August 22 and 2008 September 03, observing 29 bursts. Two of these (burst IDs {\it 080824054} and {\it 080825200}) had saturated parts, and were therefore excluded from the analysis due to the rather complicated effects saturation can have on periodograms. Following \citet{2011ApJ...739...87L}, we used only NaI detectors with a source angle smaller than $50\deg$ for each of the 24 triggered and 3 untriggered bursts. We use high-time resolution time-tagged event (TTE) data, for which time and energy for each individual photon is recorded. The data were barycentered and channels converted to the mid-energy of each energy bin. The observations were then energy-selected to include only counts between $8$ and $100$ keV. The lower limit to the energy is set by the detector response \citep{2009ApJ...702..791M}, the upper limit was found by inspecting energy-resolved light curves and finding no source counts above 100 keV (the counts were consistent with the Poisson distribution expected from counting noise). 
Burst start times and lengths (T90 durations) were taken from \citet{2011ApJ...739...87L}, and are summarized in Table 1 of that paper. We added $20\%$ of the burst duration to both ends of the burst in order to ensure that we caught the entire burst. A selection of six bursts is shown in Figure \ref{fig:burstselect}, to emphasize the diversity of burst morphologies we encounter.

\section{Results}
\begin{figure*}
\begin{center}
\includegraphics[width=16cm, height=6.0cm]{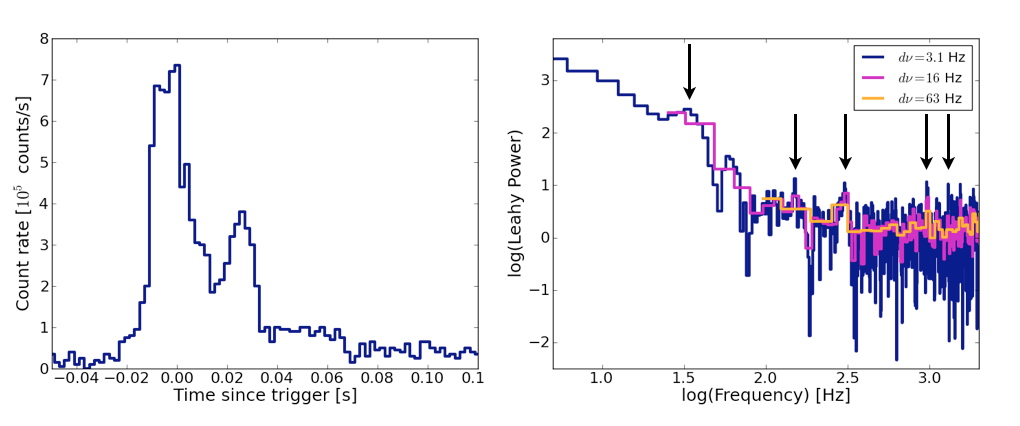}
\caption{Fermi/{\it GBM} observation of burst {\it 080823847a} from SGR J$0501+4516$. Left: light curve with a time resolution of 0.002 seconds. Structure in the burst profile is clearly visible. Right: unbinned periodogram (blue) and two examples of binned (magenta: 16 Hz binning; orange: 65 Hz binning) periodograms for this burst. There is a feature in the periodogram around 30 Hz (leftmost arrow), which is by itself not significant. However, significant features (arrows 2-5) are all at integer multiples of this frequency (within the uncertainty imposed by the frequency resolution), indicating the presence of harmonics at 150 Hz, 300 Hz, 900 Hz and 2100 Hz.}
\label{fig:080823847a_lcps}
\end{center}
\end{figure*}

We computed light curves and periodograms for all 27 bursts. In each case, we produced a light curve by binning the TTE data to a time resolution of $1/2\nu_{\mathrm{Nyquist}} = 1.22\times 10^{-4} \, \mathrm{s}$, corresponding to a Nyquist frequency of $\nu_{\mathrm{Nyquist}} = 4096 \, \mathrm{Hz}$. We chose the time resolution based on the Nyquist frequency of interest: we do not expect any signals above $4000 \, \mathrm{Hz}$ from neutron star seismic oscillations \citep{1988ApJ...325..725M}. 
We search both the unbinned periodogram as well as the same periodogram binned to integer multiples (3, 5, 7, 10, 15, 20, 30, 70, 100, 200, 300, 500 and 700) of the frequency resolution of that burst.

\subsection{Search for QPOs}

None of the 27 bursts shows periodicities of any significance in any of the unbinned periodograms. The highest data/model outlier significance is seen in burst {\it bn080823847a} (see Figure \ref{fig:080823847a_lcps} for a light curve and periodogram), with a posterior p-value $p(T_R) = 0.11 \pm 0.01$, at frequency $\nu_{\mathrm{max}} = 4057 \, \mathrm{Hz}$, well below the power required to reach the detection threshold corresponding to a posterior p-value of 5\%. However, the same burst shows significant signals from \mytilde$300$ Hz to \mytilde$2100$ Hz in the several binned periodograms (binning frequencies between 16 Hz and 1583 Hz). The most significant signals are for binning frequencies $95$ Hz and $158$ Hz at a frequency of \mytilde$2100$ Hz with a posterior p-value $< 2.0\times 10^{-5}$. Note that p-values quoted there are corrected for the number of frequencies searched, but neither for the number of bursts searched nor the number of binned spectra searched for each burst. While the former is straightforward (a simple multiplication factor of 27 for the number of bursts searched), the latter is more complicated, owing to the fact that searching different binnings for a single periodogram does not result in independent trials. The most conservative assumption is to consider them independent, including another multiplication factor equal to the number of binnings searched (here: 9). This would rule out all but the two signals with frequency bins of $95$ and $158$ Hz, which remain significant even after a correction for the number of trials. A possible interpretation of the observed signals follows in Section \ref{sec:discussion}.

\subsection{Broadband Variability}
The broadband variability observed in the bursts is not just a nuisance when searching for (quasi-) periodicities, but is of interest in its own right: it shows that something is varying in the source, although not periodically. 
Almost all bursts in the sample are well-modeled by a simple power law. The distribution of indices ranges from 1.7 to 4.3 and peaks around 2.5, which is higher than commonly seen for example in Gamma-ray bursts (see e.g. \citealt{2000ApJ...535..158B} and \citealt{2012MNRAS.422.1785G}). Only two bursts required the more complex model. While these were not the longest bursts, they had the highest fluence (except for the excluded saturated bursts), indicating a potential correlation between power spectral shape and burst fluence. A reliable characterisation of the broadband properties will be deferred to a future paper involving a larger sample of bursts.

\section{Discussion}
\label{sec:discussion}
Magnetar bursts are a potential window into the interior of neutron stars, via the oscillations measured in magnetar giant flares. Finding analogous signals in the wealth of short SGR bursts, however, poses something of a challenge. We have shown that timing analysis of astrophysical transients is a non-trivial problem. Standard Fourier techniques are insufficient for phenomena with diverse light curve morphologies, especially when involving correlated noise processes.  
Monte Carlo simulations of light curves fail to be predictive when there is no precise knowledge of the underlying burst light curve: there is a degeneracy between the overall, aperiodic burst shape, a potential red noise component, and the very thing we would like to measure: a QPO.  

In the absence of better knowledge about the emission processes in magnetar bursts, we advocate a conservative Bayesian method that models the burst light curve as a pure red noise process, at the cost that weak signals are likely missed. This is the greatest weakness of our approach. Even strong signals may be undetectable at low frequencies, where burst envelope and red noise dominate. This limitation is in part not only due to restrictions of our method, but also to the short lengths of the SGR bursts, where at these frequencies only one or two cycles may be seen in the light curve. However, at frequencies close to and above 100 Hz, sensitivities approach the white noise limit, which is strongly dependent on the number of photons from a particular burst. Thus, for a bright burst with good count statistics, sensitivities are quite constraining, down to less than 10 \%. This is comparable to what was observed in giant flares: for example, a QPO at 93 Hz, as seen in the 2004 flare, at roughly 10 \% rms amplitude \citep{2005ApJ...628L..53I, 2006ApJ...637L.117W}, should be detectable in at least the brightest bursts of our sample.

However, QPOs in SGR bursts may be less strong than in the giant flares, owing to the lower energy injected in SGR bursts, and hence more likely to be misclassified as non-detections, if their fractional rms amplitudes fall below 5 \%. 

The burst {\it 080823847a} presents an interesting case that illustrates the limits of a pure signal-processing approach to the timing analysis shown here. The nature of the significant signals is at present unclear. They are possibly harmonics of a lower-frequency signal around $30$ Hz, corresponding to a timescale of $\tau = 1/\nu = 33$ ms. This timescale roughly corresponds to the two sharp peaks seen in the burst light curve in Figure \ref{fig:080823847a_lcps} (left side). Whether we consider this to be a QPO atop a burst envelope or not cannot be answered from Fourier analysis alone; it becomes a matter of interpretation and prior knowledge. At present, without any knowledge about emission processes and the kind of light curve they produce, it is impossible to distinguish whether the two sharp peaks are indeed a heavily damped QPO, or simply a chance occurrence of red noise features, thus we choose the conservative approach and interpret the observed feature as part of a noise process. 
 
We wish to note that methods presented here, while developed with SGR bursts in mind, are by no means limited to magnetars. They are applicable in fairly general circumstances, for any light curve that is phenomenologically similar to what we observe from magnetars: highly variable, transient events with complex light curves. This includes, for example, other known transients such as gamma-ray bursts (GRBs), tidal disruption events and supernova light curves. \\

%

\begin{acknowledgments}
D.H. and A.L.W acknowledge support from a Netherlands Organization for Scientific Research (NWO) Vidi Fellowship (PI A. Watts), and would like to thank Jason Farquhar for useful discussions. C.K. and was partially supported by NASA grant NNH07ZDA001-GLAST.  E.G. acknowledges support from the Scientific and Technological Research Council of Turkey (T\"UB\"ITAK) through grant 109T755.  

\end{acknowledgments}

\bigskip 
\bibliographystyle{apsrev}
\bibliography{sgr0501_paper_references}

\begin{thebibliography}{29}
\expandafter\ifx\csname natexlab\endcsname\relax\def\natexlab#1{#1}\fi
\expandafter\ifx\csname bibnamefont\endcsname\relax
  \def\bibnamefont#1{#1}\fi
\expandafter\ifx\csname bibfnamefont\endcsname\relax
  \def\bibfnamefont#1{#1}\fi
\expandafter\ifx\csname citenamefont\endcsname\relax
  \def\citenamefont#1{#1}\fi
\expandafter\ifx\csname url\endcsname\relax
  \def\url#1{\texttt{#1}}\fi
\expandafter\ifx\csname urlprefix\endcsname\relax\def\urlprefix{URL }\fi
\providecommand{\bibinfo}[2]{#2}
\providecommand{\eprint}[2][]{\url{#2}}

\bibitem[{\citenamefont{{Beloborodov} et~al.}(2000)\citenamefont{{Beloborodov},
  {Stern}, and {Svensson}}}]{2000ApJ...535..158B}
\bibinfo{author}{\bibfnamefont{A.~M.} \bibnamefont{{Beloborodov}}},
  \bibinfo{author}{\bibfnamefont{B.~E.} \bibnamefont{{Stern}}},
  \bibnamefont{and}
  \bibinfo{author}{\bibfnamefont{R.}~\bibnamefont{{Svensson}}},
  \bibinfo{journal}{Astrophysical Journal} \textbf{\bibinfo{volume}{535}},
  \bibinfo{pages}{158} (\bibinfo{year}{2000}).

\bibitem[{\citenamefont{{Duncan}}(1998)}]{1998ApJ...498L..45D}
\bibinfo{author}{\bibfnamefont{R.~C.} \bibnamefont{{Duncan}}},
  \bibinfo{journal}{Astrophysical Journal Letters}
  \textbf{\bibinfo{volume}{498}}, \bibinfo{pages}{L45} (\bibinfo{year}{1998}),
  \eprint{arXiv:astro-ph/9803060}.

\bibitem[{\citenamefont{{El-Mezeini} and
  {Ibrahim}}(2010)}]{2010ApJ...721L.121E}
\bibinfo{author}{\bibfnamefont{A.~M.} \bibnamefont{{El-Mezeini}}}
  \bibnamefont{and} \bibinfo{author}{\bibfnamefont{A.~I.}
  \bibnamefont{{Ibrahim}}}, \bibinfo{journal}{Astrophysical Journal Letters}
  \textbf{\bibinfo{volume}{721}}, \bibinfo{pages}{L121} (\bibinfo{year}{2010}),
  \eprint{1008.3870}.

\bibitem[{\citenamefont{{Foreman-Mackey}
  et~al.}(2012)\citenamefont{{Foreman-Mackey}, {Hogg}, {Lang}, and
  {Goodman}}}]{2012arXiv1202.3665F}
\bibinfo{author}{\bibfnamefont{D.}~\bibnamefont{{Foreman-Mackey}}},
  \bibinfo{author}{\bibfnamefont{D.~W.} \bibnamefont{{Hogg}}},
  \bibinfo{author}{\bibfnamefont{D.}~\bibnamefont{{Lang}}}, \bibnamefont{and}
  \bibinfo{author}{\bibfnamefont{J.}~\bibnamefont{{Goodman}}},
  \bibinfo{journal}{ArXiv e-prints} p. \bibinfo{pages}{1202.3665}
  (\bibinfo{year}{2012}).



\bibitem[{\citenamefont{{Fox} et~al.}(2001)\citenamefont{{Fox}, {Lewin},
  {Rutledge}, {Morgan}, {Guerriero}, {Bildsten}, {van der Klis}, {van
  Paradijs}, {Moore}, {Dotani} et~al.}}]{2001MNRAS.321..776F}
\bibinfo{author}{\bibfnamefont{D.~W.} \bibnamefont{{Fox}}},
  \bibinfo{author}{\bibfnamefont{W.~H.~G.} \bibnamefont{{Lewin}}},
  \bibinfo{author}{\bibfnamefont{R.~E.} \bibnamefont{{Rutledge}}},
  \bibinfo{author}{\bibfnamefont{E.~H.} \bibnamefont{{Morgan}}},
  \bibinfo{author}{\bibfnamefont{R.}~\bibnamefont{{Guerriero}}},
  \bibinfo{author}{\bibfnamefont{L.}~\bibnamefont{{Bildsten}}},
  \bibinfo{author}{\bibfnamefont{M.}~\bibnamefont{{van der Klis}}},
  \bibinfo{author}{\bibfnamefont{J.}~\bibnamefont{{van Paradijs}}},
  \bibinfo{author}{\bibfnamefont{C.~B.} \bibnamefont{{Moore}}},
  \bibinfo{author}{\bibfnamefont{T.}~\bibnamefont{{Dotani}}},
  \bibnamefont{et~al.}, \bibinfo{journal}{Monthly Notices of the Royal
  Astronomical Society} \textbf{\bibinfo{volume}{321}}, \bibinfo{pages}{776}
  (\bibinfo{year}{2001}), \eprint{arXiv:astro-ph/0009224}.

\bibitem[{\citenamefont{{G{\"o}{\u g}{\"u}{\c s} }
  et~al.}(1999)\citenamefont{{G{\"o}{\u g}{\"u}{\c s} }, {Woods},
  {Kouveliotou}, {van Paradijs}, {Briggs}, {Duncan}, and
  {Thompson}}}]{1999ApJ...526L..93G}
\bibinfo{author}{\bibfnamefont{E.}~\bibnamefont{{G{\"o}{\u g}{\"u}{\c s} }}},
  \bibinfo{author}{\bibfnamefont{P.~M.} \bibnamefont{{Woods}}},
  \bibinfo{author}{\bibfnamefont{C.}~\bibnamefont{{Kouveliotou}}},
  \bibinfo{author}{\bibfnamefont{J.}~\bibnamefont{{van Paradijs}}},
  \bibinfo{author}{\bibfnamefont{M.~S.} \bibnamefont{{Briggs}}},
  \bibinfo{author}{\bibfnamefont{R.~C.} \bibnamefont{{Duncan}}},
  \bibnamefont{and}
  \bibinfo{author}{\bibfnamefont{C.}~\bibnamefont{{Thompson}}},
  \bibinfo{journal}{Astrophysical Journal Letters}
  \textbf{\bibinfo{volume}{526}}, \bibinfo{pages}{L93} (\bibinfo{year}{1999}),
  \eprint{arXiv:astro-ph/9910062}.

\bibitem[{\citenamefont{{G{\"o}{\u g}{\"u}{\c s}}
  et~al.}(2000)\citenamefont{{G{\"o}{\u g}{\"u}{\c s}}, {Woods}, {Kouveliotou},
  {van Paradijs}, {Briggs}, {Duncan}, and {Thompson}}}]{2000ApJ...532L.121G}
\bibinfo{author}{\bibfnamefont{E.}~\bibnamefont{{G{\"o}{\u g}{\"u}{\c s}}}},
  \bibinfo{author}{\bibfnamefont{P.~M.} \bibnamefont{{Woods}}},
  \bibinfo{author}{\bibfnamefont{C.}~\bibnamefont{{Kouveliotou}}},
  \bibinfo{author}{\bibfnamefont{J.}~\bibnamefont{{van Paradijs}}},
  \bibinfo{author}{\bibfnamefont{M.~S.} \bibnamefont{{Briggs}}},
  \bibinfo{author}{\bibfnamefont{R.~C.} \bibnamefont{{Duncan}}},
  \bibnamefont{and}
  \bibinfo{author}{\bibfnamefont{C.}~\bibnamefont{{Thompson}}},
  \bibinfo{journal}{Astrophysical Journal Letters}
  \textbf{\bibinfo{volume}{532}}, \bibinfo{pages}{L121} (\bibinfo{year}{2000}),
  \eprint{arXiv:astro-ph/0002181}.

\bibitem[{\citenamefont{{Guidorzi}}(2011)}]{2011MNRAS.415.3561G}
\bibinfo{author}{\bibfnamefont{C.}~\bibnamefont{{Guidorzi}}},
  \bibinfo{journal}{Monthly Notices of the Royal Astronomical Society}
  \textbf{\bibinfo{volume}{415}}, \bibinfo{pages}{3561} (\bibinfo{year}{2011}),
  \eprint{1104.5308}.

\bibitem[{\citenamefont{{Guidorzi} et~al.}(2012)\citenamefont{{Guidorzi},
  {Margutti}, {Amati}, {Campana}, {Orlandini}, {Romano}, {Stamatikos}, and
  {Tagliaferri}}}]{2012MNRAS.422.1785G}
\bibinfo{author}{\bibfnamefont{C.}~\bibnamefont{{Guidorzi}}},
  \bibinfo{author}{\bibfnamefont{R.}~\bibnamefont{{Margutti}}},
  \bibinfo{author}{\bibfnamefont{L.}~\bibnamefont{{Amati}}},
  \bibinfo{author}{\bibfnamefont{S.}~\bibnamefont{{Campana}}},
  \bibinfo{author}{\bibfnamefont{M.}~\bibnamefont{{Orlandini}}},
  \bibinfo{author}{\bibfnamefont{P.}~\bibnamefont{{Romano}}},
  \bibinfo{author}{\bibfnamefont{M.}~\bibnamefont{{Stamatikos}}},
  \bibnamefont{and}
  \bibinfo{author}{\bibfnamefont{G.}~\bibnamefont{{Tagliaferri}}},
  \bibinfo{journal}{Monthly Notices of the Royal Astronomical Society}
  \textbf{\bibinfo{volume}{422}}, \bibinfo{pages}{1785} (\bibinfo{year}{2012}),
  \eprint{1202.3443}.


\bibitem[{\citenamefont{{Huppenkothen}
  et~al.}(2013)\citenamefont{{Huppenkothen}, {Watts}, {Uttley}, {van der
  Horst}, {van der Klis}, {Kouveliotou}, {Gogus}, {Granot}, {Vaughan}, and
  {Finger}}}]{sgr0501_timing}
\bibinfo{author}{\bibfnamefont{D.}~\bibnamefont{{Huppenkothen}}},
  \bibinfo{author}{\bibfnamefont{A.~L.} \bibnamefont{{Watts}}},
  \bibinfo{author}{\bibfnamefont{P.}~\bibnamefont{{Uttley}}},
  \bibinfo{author}{\bibfnamefont{A.~J.} \bibnamefont{{van der Horst}}},
  \bibinfo{author}{\bibfnamefont{M.}~\bibnamefont{{van der Klis}}},
  \bibinfo{author}{\bibfnamefont{C.}~\bibnamefont{{Kouveliotou}}},
  \bibinfo{author}{\bibfnamefont{E.}~\bibnamefont{{Gogus}}},
  \bibinfo{author}{\bibfnamefont{J.}~\bibnamefont{{Granot}}},
  \bibinfo{author}{\bibfnamefont{S.}~\bibnamefont{{Vaughan}}},
  \bibnamefont{and} \bibinfo{author}{\bibfnamefont{M.~H.}
  \bibnamefont{{Finger}}}, \bibinfo{journal}{Astrophysical Journal, in press}
  (\bibinfo{year}{2013}), \eprint{1212.1011}.

\bibitem[{\citenamefont{{Hurley} et~al.}(1999)\citenamefont{{Hurley}, {Cline},
  {Mazets}, {Barthelmy}, {Butterworth}, {Marshall}, {Palmer}, {Aptekar},
  {Golenetskii}, {Il'Inskii} et~al.}}]{1999Natur.397...41H}
\bibinfo{author}{\bibfnamefont{K.}~\bibnamefont{{Hurley}}},
  \bibinfo{author}{\bibfnamefont{T.}~\bibnamefont{{Cline}}},
  \bibinfo{author}{\bibfnamefont{E.}~\bibnamefont{{Mazets}}},
  \bibinfo{author}{\bibfnamefont{S.}~\bibnamefont{{Barthelmy}}},
  \bibinfo{author}{\bibfnamefont{P.}~\bibnamefont{{Butterworth}}},
  \bibinfo{author}{\bibfnamefont{F.}~\bibnamefont{{Marshall}}},
  \bibinfo{author}{\bibfnamefont{D.}~\bibnamefont{{Palmer}}},
  \bibinfo{author}{\bibfnamefont{R.}~\bibnamefont{{Aptekar}}},
  \bibinfo{author}{\bibfnamefont{S.}~\bibnamefont{{Golenetskii}}},
  \bibinfo{author}{\bibfnamefont{V.}~\bibnamefont{{Il'Inskii}}},
  \bibnamefont{et~al.}, \bibinfo{journal}{Nature}
  \textbf{\bibinfo{volume}{397}}, \bibinfo{pages}{41} (\bibinfo{year}{1999}),
  \eprint{arXiv:astro-ph/9811443}.

\bibitem[{\citenamefont{{Israel} et~al.}(2005)\citenamefont{{Israel},
  {Belloni}, {Stella}, {Rephaeli}, {Gruber}, {Casella}, {Dall'Osso}, {Rea},
  {Persic}, and {Rothschild}}}]{2005ApJ...628L..53I}
\bibinfo{author}{\bibfnamefont{G.~L.} \bibnamefont{{Israel}}},
  \bibinfo{author}{\bibfnamefont{T.}~\bibnamefont{{Belloni}}},
  \bibinfo{author}{\bibfnamefont{L.}~\bibnamefont{{Stella}}},
  \bibinfo{author}{\bibfnamefont{Y.}~\bibnamefont{{Rephaeli}}},
  \bibinfo{author}{\bibfnamefont{D.~E.} \bibnamefont{{Gruber}}},
  \bibinfo{author}{\bibfnamefont{P.}~\bibnamefont{{Casella}}},
  \bibinfo{author}{\bibfnamefont{S.}~\bibnamefont{{Dall'Osso}}},
  \bibinfo{author}{\bibfnamefont{N.}~\bibnamefont{{Rea}}},
  \bibinfo{author}{\bibfnamefont{M.}~\bibnamefont{{Persic}}}, \bibnamefont{and}
  \bibinfo{author}{\bibfnamefont{R.~E.} \bibnamefont{{Rothschild}}},
  \bibinfo{journal}{Astrophysical Journal Letters}
  \textbf{\bibinfo{volume}{628}}, \bibinfo{pages}{L53} (\bibinfo{year}{2005}),
  \eprint{arXiv:astro-ph/0505255}.


\bibitem[{\citenamefont{{Kaneko} et~al.}(2010)\citenamefont{{Kaneko},
  {G{\"o}{\u g}{\"u}{\c s}}, {Kouveliotou}, {Granot}, {Ramirez-Ruiz}, {van der
  Horst}, {Watts}, {Finger}, {Gehrels}, {Pe'er} et~al.}}]{Kaneko10}
\bibinfo{author}{\bibfnamefont{Y.}~\bibnamefont{{Kaneko}}},
  \bibinfo{author}{\bibfnamefont{E.}~\bibnamefont{{G{\"o}{\u g}{\"u}{\c s}}}},
  \bibinfo{author}{\bibfnamefont{C.}~\bibnamefont{{Kouveliotou}}},
  \bibinfo{author}{\bibfnamefont{J.}~\bibnamefont{{Granot}}},
  \bibinfo{author}{\bibfnamefont{E.}~\bibnamefont{{Ramirez-Ruiz}}},
  \bibinfo{author}{\bibfnamefont{A.~J.} \bibnamefont{{van der Horst}}},
  \bibinfo{author}{\bibfnamefont{A.~L.} \bibnamefont{{Watts}}},
  \bibinfo{author}{\bibfnamefont{M.~H.} \bibnamefont{{Finger}}},
  \bibinfo{author}{\bibfnamefont{N.}~\bibnamefont{{Gehrels}}},
  \bibinfo{author}{\bibfnamefont{A.}~\bibnamefont{{Pe'er}}},
  \bibnamefont{et~al.}, \bibinfo{journal}{Astrophysical Journal}
  \textbf{\bibinfo{volume}{710}}, \bibinfo{pages}{1335} (\bibinfo{year}{2010}),
  \eprint{0911.4636}.


\bibitem[{\citenamefont{{Kouveliotou} et~al.}(1998)\citenamefont{{Kouveliotou},
  {Dieters}, {Strohmayer}, {van Paradijs}, {Fishman}, {Meegan}, {Hurley},
  {Kommers}, {Smith}, {Frail} et~al.}}]{1998Natur.393..235K}
\bibinfo{author}{\bibfnamefont{C.}~\bibnamefont{{Kouveliotou}}},
  \bibinfo{author}{\bibfnamefont{S.}~\bibnamefont{{Dieters}}},
  \bibinfo{author}{\bibfnamefont{T.}~\bibnamefont{{Strohmayer}}},
  \bibinfo{author}{\bibfnamefont{J.}~\bibnamefont{{van Paradijs}}},
  \bibinfo{author}{\bibfnamefont{G.~J.} \bibnamefont{{Fishman}}},
  \bibinfo{author}{\bibfnamefont{C.~A.} \bibnamefont{{Meegan}}},
  \bibinfo{author}{\bibfnamefont{K.}~\bibnamefont{{Hurley}}},
  \bibinfo{author}{\bibfnamefont{J.}~\bibnamefont{{Kommers}}},
  \bibinfo{author}{\bibfnamefont{I.}~\bibnamefont{{Smith}}},
  \bibinfo{author}{\bibfnamefont{D.}~\bibnamefont{{Frail}}},
  \bibnamefont{et~al.}, \bibinfo{journal}{Astrophysical JournalNature}
  \textbf{\bibinfo{volume}{393}}, \bibinfo{pages}{235} (\bibinfo{year}{1998}).

\bibitem[{\citenamefont{{Kouveliotou} et~al.}(1999)\citenamefont{{Kouveliotou},
  {Strohmayer}, {Hurley}, {van Paradijs}, {Finger}, {Dieters}, {Woods},
  {Thompson}, and {Duncan}}}]{1999ApJ...510L.115K}
\bibinfo{author}{\bibfnamefont{C.}~\bibnamefont{{Kouveliotou}}},
  \bibinfo{author}{\bibfnamefont{T.}~\bibnamefont{{Strohmayer}}},
  \bibinfo{author}{\bibfnamefont{K.}~\bibnamefont{{Hurley}}},
  \bibinfo{author}{\bibfnamefont{J.}~\bibnamefont{{van Paradijs}}},
  \bibinfo{author}{\bibfnamefont{M.~H.} \bibnamefont{{Finger}}},
  \bibinfo{author}{\bibfnamefont{S.}~\bibnamefont{{Dieters}}},
  \bibinfo{author}{\bibfnamefont{P.}~\bibnamefont{{Woods}}},
  \bibinfo{author}{\bibfnamefont{C.}~\bibnamefont{{Thompson}}},
  \bibnamefont{and} \bibinfo{author}{\bibfnamefont{R.~C.}
  \bibnamefont{{Duncan}}}, \bibinfo{journal}{Astrophysical Journal Letters}
  \textbf{\bibinfo{volume}{510}}, \bibinfo{pages}{L115} (\bibinfo{year}{1999}),
  \eprint{arXiv:astro-ph/9809140}.


\bibitem[{\citenamefont{{Lattimer} and {Prakash}}(2007)}]{Lattimer07}
\bibinfo{author}{\bibfnamefont{J.~M.} \bibnamefont{{Lattimer}}}
  \bibnamefont{and}
  \bibinfo{author}{\bibfnamefont{M.}~\bibnamefont{{Prakash}}},
  \bibinfo{journal}{Physics Reports} \textbf{\bibinfo{volume}{442}},
  \bibinfo{pages}{109} (\bibinfo{year}{2007}), \eprint{arXiv:astro-ph/0612440}.

\bibitem[{\citenamefont{{Lin} et~al.}(2011)\citenamefont{{Lin}, {Kouveliotou},
  {Baring}, {van der Horst}, {Guiriec}, {Woods}, {G{\"o}{\u g}{\"u}{\c s}},
  {Kaneko}, {Scargle}, {Granot} et~al.}}]{2011ApJ...739...87L}
\bibinfo{author}{\bibfnamefont{L.}~\bibnamefont{{Lin}}},
  \bibinfo{author}{\bibfnamefont{C.}~\bibnamefont{{Kouveliotou}}},
  \bibinfo{author}{\bibfnamefont{M.~G.} \bibnamefont{{Baring}}},
  \bibinfo{author}{\bibfnamefont{A.~J.} \bibnamefont{{van der Horst}}},
  \bibinfo{author}{\bibfnamefont{S.}~\bibnamefont{{Guiriec}}},
  \bibinfo{author}{\bibfnamefont{P.~M.} \bibnamefont{{Woods}}},
  \bibinfo{author}{\bibfnamefont{E.}~\bibnamefont{{G{\"o}{\u g}{\"u}{\c s}}}},
  \bibinfo{author}{\bibfnamefont{Y.}~\bibnamefont{{Kaneko}}},
  \bibinfo{author}{\bibfnamefont{J.}~\bibnamefont{{Scargle}}},
  \bibinfo{author}{\bibfnamefont{J.}~\bibnamefont{{Granot}}},
  \bibnamefont{et~al.}, \bibinfo{journal}{Astrophysical Journal}
  \textbf{\bibinfo{volume}{739}}, \bibinfo{eid}{87} (\bibinfo{year}{2011}).

\bibitem[{\citenamefont{{McDermott} et~al.}(1988)\citenamefont{{McDermott},
  {van Horn}, and {Hansen}}}]{1988ApJ...325..725M}
\bibinfo{author}{\bibfnamefont{P.~N.} \bibnamefont{{McDermott}}},
  \bibinfo{author}{\bibfnamefont{H.~M.} \bibnamefont{{van Horn}}},
  \bibnamefont{and} \bibinfo{author}{\bibfnamefont{C.~J.}
  \bibnamefont{{Hansen}}}, \bibinfo{journal}{Astrophysical Journal}
  \textbf{\bibinfo{volume}{325}}, \bibinfo{pages}{725} (\bibinfo{year}{1988}).

\bibitem[{\citenamefont{{Meegan} et~al.}(2009)\citenamefont{{Meegan}, {Lichti},
  {Bhat}, {Bissaldi}, {Briggs}, {Connaughton}, {Diehl}, {Fishman}, {Greiner},
  {Hoover} et~al.}}]{2009ApJ...702..791M}
\bibinfo{author}{\bibfnamefont{C.}~\bibnamefont{{Meegan}}},
  \bibinfo{author}{\bibfnamefont{G.}~\bibnamefont{{Lichti}}},
  \bibinfo{author}{\bibfnamefont{P.~N.} \bibnamefont{{Bhat}}},
  \bibinfo{author}{\bibfnamefont{E.}~\bibnamefont{{Bissaldi}}},
  \bibinfo{author}{\bibfnamefont{M.~S.} \bibnamefont{{Briggs}}},
  \bibinfo{author}{\bibfnamefont{V.}~\bibnamefont{{Connaughton}}},
  \bibinfo{author}{\bibfnamefont{R.}~\bibnamefont{{Diehl}}},
  \bibinfo{author}{\bibfnamefont{G.}~\bibnamefont{{Fishman}}},
  \bibinfo{author}{\bibfnamefont{J.}~\bibnamefont{{Greiner}}},
  \bibinfo{author}{\bibfnamefont{A.~S.} \bibnamefont{{Hoover}}},
  \bibnamefont{et~al.}, \bibinfo{journal}{Astrophysical Journal}
  \textbf{\bibinfo{volume}{702}}, \bibinfo{pages}{791} (\bibinfo{year}{2009}),
  \eprint{0908.0450}.



\bibitem[{\citenamefont{{Mereghetti}}(2008)}]{mereghetti08}
\bibinfo{author}{\bibfnamefont{S.}~\bibnamefont{{Mereghetti}}},
  \bibinfo{journal}{The Astronomy and Astrophysics Review}
  \textbf{\bibinfo{volume}{15}}, \bibinfo{pages}{225} (\bibinfo{year}{2008}),
  \eprint{0804.0250}.


\bibitem[{\citenamefont{{Palmer} et~al.}(2005)\citenamefont{{Palmer},
  {Barthelmy}, {Gehrels}, {Kippen}, {Cayton}, {Kouveliotou}, {Eichler},
  {Wijers}, {Woods}, {Granot} et~al.}}]{2005Natur.434.1107P}
\bibinfo{author}{\bibfnamefont{D.~M.} \bibnamefont{{Palmer}}},
  \bibinfo{author}{\bibfnamefont{S.}~\bibnamefont{{Barthelmy}}},
  \bibinfo{author}{\bibfnamefont{N.}~\bibnamefont{{Gehrels}}},
  \bibinfo{author}{\bibfnamefont{R.~M.} \bibnamefont{{Kippen}}},
  \bibinfo{author}{\bibfnamefont{T.}~\bibnamefont{{Cayton}}},
  \bibinfo{author}{\bibfnamefont{C.}~\bibnamefont{{Kouveliotou}}},
  \bibinfo{author}{\bibfnamefont{D.}~\bibnamefont{{Eichler}}},
  \bibinfo{author}{\bibfnamefont{R.~A.~M.~J.} \bibnamefont{{Wijers}}},
  \bibinfo{author}{\bibfnamefont{P.~M.} \bibnamefont{{Woods}}},
  \bibinfo{author}{\bibfnamefont{J.}~\bibnamefont{{Granot}}},
  \bibnamefont{et~al.}, \bibinfo{journal}{Nature}
  \textbf{\bibinfo{volume}{434}}, \bibinfo{pages}{1107} (\bibinfo{year}{2005}),
  \eprint{arXiv:astro-ph/0503030}.

\bibitem[{\citenamefont{{Strohmayer} and {Watts}}(2005)}]{2005ApJ...632L.111S}
\bibinfo{author}{\bibfnamefont{T.~E.} \bibnamefont{{Strohmayer}}}
  \bibnamefont{and} \bibinfo{author}{\bibfnamefont{A.~L.}
  \bibnamefont{{Watts}}}, \bibinfo{journal}{Astrophysical Journal Letters}
  \textbf{\bibinfo{volume}{632}}, \bibinfo{pages}{L111} (\bibinfo{year}{2005}),
  \eprint{arXiv:astro-ph/0508206}.

\bibitem[{\citenamefont{{Strohmayer} and {Watts}}(2006)}]{2006ApJ...653..593S}
\bibinfo{author}{\bibfnamefont{T.~E.} \bibnamefont{{Strohmayer}}}
  \bibnamefont{and} \bibinfo{author}{\bibfnamefont{A.~L.}
  \bibnamefont{{Watts}}}, \bibinfo{journal}{Astrophysical Journal}
  \textbf{\bibinfo{volume}{653}}, \bibinfo{pages}{593} (\bibinfo{year}{2006}),
  \eprint{arXiv:astro-ph/0608463}.



\bibitem[{\citenamefont{{Thompson} and {Duncan}}(1995)}]{1995MNRAS.275..255T}
\bibinfo{author}{\bibfnamefont{C.}~\bibnamefont{{Thompson}}} \bibnamefont{and}
  \bibinfo{author}{\bibfnamefont{R.~C.} \bibnamefont{{Duncan}}},
  \bibinfo{journal}{Monthly Notices of the Royal Astronomical Society}
  \textbf{\bibinfo{volume}{275}}, \bibinfo{pages}{255} (\bibinfo{year}{1995}).

\bibitem[{\citenamefont{{Vaughan}}(2010)}]{vaughan2010}
\bibinfo{author}{\bibfnamefont{S.}~\bibnamefont{{Vaughan}}},
  \bibinfo{journal}{Monthly Notices of the Royal Astronomical Society}
  \textbf{\bibinfo{volume}{402}}, \bibinfo{pages}{307} (\bibinfo{year}{2010}),
  \eprint{0910.2706}.

\bibitem[{\citenamefont{{Watts} and {Strohmayer}}(2006)}]{2006ApJ...637L.117W}
\bibinfo{author}{\bibfnamefont{A.~L.} \bibnamefont{{Watts}}} \bibnamefont{and}
  \bibinfo{author}{\bibfnamefont{T.~E.} \bibnamefont{{Strohmayer}}},
  \bibinfo{journal}{Astrophysical Journal Letters}
  \textbf{\bibinfo{volume}{637}}, \bibinfo{pages}{L117} (\bibinfo{year}{2006}),
  \eprint{arXiv:astro-ph/0512630}.


\bibitem[{\citenamefont{{Watts}}(2012)}]{Watts11}
\bibinfo{author}{\bibfnamefont{A.~L.} \bibnamefont{{Watts}}}, in
  \emph{\bibinfo{booktitle}{Neutron Star Crust}}, edited by
  \bibinfo{editor}{\bibfnamefont{C.~A.} \bibnamefont{{Bertulani}}}
  \bibnamefont{and}
  \bibinfo{editor}{\bibfnamefont{J.}~\bibnamefont{{Piekarewicz}}}
  (\bibinfo{publisher}{Nova Science Pub Inc}, \bibinfo{year}{2012}), pp.
  \bibinfo{pages}{265--280}
 \eprint{arXiv:astro-ph/1111.0514}.


\bibitem[{\citenamefont{{Woods} et~al.}(2001)\citenamefont{{Woods},
  {Kouveliotou}, {G{\"o}{\u g}{\"u}{\c s}}, {Finger}, {Swank}, {Smith},
  {Hurley}, and {Thompson}}}]{2001ApJ...552..748W}
\bibinfo{author}{\bibfnamefont{P.~M.} \bibnamefont{{Woods}}},
  \bibinfo{author}{\bibfnamefont{C.}~\bibnamefont{{Kouveliotou}}},
  \bibinfo{author}{\bibfnamefont{E.}~\bibnamefont{{G{\"o}{\u g}{\"u}{\c s}}}},
  \bibinfo{author}{\bibfnamefont{M.~H.} \bibnamefont{{Finger}}},
  \bibinfo{author}{\bibfnamefont{J.}~\bibnamefont{{Swank}}},
  \bibinfo{author}{\bibfnamefont{D.~A.} \bibnamefont{{Smith}}},
  \bibinfo{author}{\bibfnamefont{K.}~\bibnamefont{{Hurley}}}, \bibnamefont{and}
  \bibinfo{author}{\bibfnamefont{C.}~\bibnamefont{{Thompson}}},
  \bibinfo{journal}{Astrophysical Journal} \textbf{\bibinfo{volume}{552}},
  \bibinfo{pages}{748} (\bibinfo{year}{2001}), \eprint{arXiv:astro-ph/0101045}.

\bibitem[{\citenamefont{{Woods} and {Thompson}}(2006)}]{woods06}
\bibinfo{author}{\bibfnamefont{P.~M.} \bibnamefont{{Woods}}} \bibnamefont{and}
  \bibinfo{author}{\bibfnamefont{C.}~\bibnamefont{{Thompson}}}, in
  \emph{\bibinfo{booktitle}{Compact stellar X-ray sources}}, edited by
  \bibinfo{editor}{\bibfnamefont{W.~H.~G.} \bibnamefont{{Lewin}}}
  \bibnamefont{and} \bibinfo{editor}{\bibfnamefont{M.}~\bibnamefont{{van der
  Klis}}} (\bibinfo{publisher}{Cambridge University Press},
  \bibinfo{year}{2006}), pp. \bibinfo{pages}{547--586}.

























\end{thebibliography}




%

\end{document}